\documentclass[12pt]{article}
\usepackage{amssymb,amsmath}
\begin{document}
\begin{flushright}
\vspace{ 1mm } hep-th/0211088
\\
FIAN/TD/17/02\\
Nov. 2002\\
\end{flushright}
\vspace{10mm}
\begin{center}

{\Large \bf Equivalent sets of solutions of the Dirac equation with
 a constant electric field}\\

 \vspace{10mm} A.I.Nikishov ${}^\dag$\\

\vspace{3mm}{$\dag$\it I.E.Tamm Department of Theoretical Physics,
              Lebedev Physical Institute,\\
 Leninsky Prospect 53, 119991, Moscow, Russia\\
             e-mail: nikishov@lpi.ru}

    \vspace{2mm}

    \end{center}
        \begin{abstract}
 Two families of sets, nonstationary and stationary, are obtained. 
Each   nonstationary
set $\psi_{p_v}$ consists of the solutions with the quantum number
 $p_v=p^0v-p_3.$
It can be obtained from the nonstationary set $\psi_{p_3}$ with quantum
 number $p_3$ by a
 boost  along $x_3$-axis ( along the direction of the electric field) with
velocity $-v$. 
Similarly, any stationary set of solutions characterized by a quantum
 number $p_s=p^0-sp_3$  can be
obtained from stationary solutions with quantum number $p^0$ by the same
boost with velocity $-s$. All these sets are equivalent and the classification
(i.e. ascribing the frequency sign and in-, out- indexes) in any set is
determined by the classification in $\psi_{p_3}$-set , where it is beyond
doubt.
\end{abstract}

\section{Introduction}

The problems connected with Klein paradox and pair production by an
external field were considered in [1-5].
The choice of the in- and out- solutions of the Klein-Gordon and Dirac equations
with one-dimensional potential
in [1-2] (see also [3]) disagrees with
 that in [4-5]. In this connection we studied in [6] the families of solutions 
 of the Klein-Gordon equation  with constant electric field and showed that
 our choice is correct. The existence of families of solutions and their
 properties are of interest by itself and here we show that the similarity
 of scalar and spinor cases is indeed very close.

 \section{ Nonstationary sets}

We use the metric
$$
\eta_{\mu\nu}={\rm diag(-1,1,1,1)},  \eqno(1)
$$
and denote the electron charge $e'=-e=-|e|$. First we show how to obtain
the nonstationary solutions characterized by the set of quantum numbers
$n=(r, p_1, p_2, p_v=p^0v-p_3)$; $r=1, 2$ indicates spin state. 
For this we take vector-potential as
$$
A_{\mu}=-\delta_{\mu3}Ex_v,\quad x_v=t-vx_3,\quad 0\le v <1.\eqno(2)
$$
The squared Dirac equation 
$$
(\Pi^2+m^2+g)F=0,\quad
 \Pi_{\mu}=-i\frac{\partial}{\partial x^{\mu}}+eA_{\mu}    \eqno(3)
$$
differs from the Klein-Gordon equation only by the presence of matrix $g$,
which we take in the spinor representation 
$$
\quad g=-ieE\alpha_3=-ieE\left(\begin{array}{cc}
\sigma_3&0\\
0&-\sigma_3
\end{array}\right), \quad \sigma_3=\left(\begin{array}{cc}
1&0\\
0&-1
\end{array}\right).                                           \eqno(4)
$$
As $g$ is diagonal in this representation, we see that the solution $F$
has the form
$$
F=\exp[i\vartheta_v]{\rm diag}(\phi_1, \phi_2, \phi_2, \phi_1),       \eqno(5)
$$                                  
where $\vartheta_v$ is (see eq.(18) in [6] with $e\to-e$)
$$
\vartheta_v= \frac{p_v(tv-x_3)}{1-v^2}+\frac{eEvx_v^2}{2(1-v^2)}. \eqno(6)
$$
The motion along $x_1$, $x_2$ remains free and we drop the corresponding 
factor $\exp[ip_1x_1+ip_2x_2]$ for brevity.

It follows from (3) and (4) that $\phi_1, \phi_2$ can be obtained from the
solutions of the Klein-Gordon equation by substitutions
$$
m^2\to m^2\mp ieE.     \eqno(7)
$$

We denote the positive-frequency in- (out-) solution of eq. (3) as ${}_+F$ 
 (${}^+F$) and similarly for negative-frequency solutions. Now we are in
 a position to find the solutions of this equation. First we note that
 the corresponding solutions of the Klein-Gordon equation are
 (see eq. (22) in [6])
$$
{}_{\pm}\psi_{p_v}=e^{i\vartheta_v}
D_{\pm i\varkappa-\frac12}(-e^{\mp\frac{\pi}{4}}T_v),\;
{}^{\pm}\psi_{p_v}=e^{i\vartheta_v}
D_{\mp i\varkappa-\frac12}(e^{\pm i\frac{\pi}{4}}T_v),
$$
  $$
T_v=\sqrt{\frac{2}{eE(1-v^2)}}(p_v+eEx_v),
\quad  \varkappa=\frac{m^2+p_1+p_2}{2eE}.                \eqno(8)
$$
Applying the substitution (7), e. g., to
 ${}^+\psi_{p_v}=e^{i\vartheta_v}
D_{- i\varkappa-\frac12}(e^{i\frac{\pi}{4}}T_v),$
we find 
$$
{}^+F=e^{i\vartheta_v}{\rm diag}\left(D_{-i\varkappa-1}(\chi_v),
 D_{-i\varkappa}(\chi_v),D_{-i\varkappa}(\chi_v),D_{-i\varkappa-1}(\chi_v)
  \right),\quad
  \chi_v=e^{\frac{i\pi}{4}}T_v. \eqno(9)
 $$
  The solutions of the Dirac equation are obtained as the columns of the matrix
$$
(m-i\hat\Pi)F=\left(\begin{array}{cccc}
m&0&\Pi^0+\Pi_3&p_1-ip_2\\
0&m&p_1+ip_2&\Pi^0-\Pi_3\\
\Pi^0-\Pi_3&-p_1+ip_2&m&0\\
-p_1-ip_2&\Pi^0+\Pi_3&0&m
\end{array}\right)F.                                              \eqno(10)
$$            
Here
$$
\Pi^0=i\frac{\partial}{\partial t},\quad \Pi_3=-i\frac{\partial}{\partial x_3}
-e'A_3=-i\frac{\partial}{\partial x_3}-eEx_v.                     \eqno(11)
$$
It is easy to verify that
$$
(\Pi^0\pm\Pi_3)e^{i\vartheta_v}\phi(\chi_v)=-e^{i\vartheta_v-\frac{i\pi}4}
\sqrt{\frac{2eE(1\pm v)}{1\mp v}}\left(\frac{\partial}{\partial\chi_v}
\pm\frac{\chi_v}2 \right)\phi(\chi_v).           \eqno(12)
$$

We use the fourth (the first) column in  matrix (10) to obtain the first
 (the second) spin state ${}^+\psi_{1p_v}$  (${}^+\psi_{2p_v}$).
 Taking into account also the first  relation in 
 $$
 \left(\frac d{dz}-\frac z2  \right) D_{\nu}(z)=-D_{\nu+1}(z),\quad
 \left(\frac d{dz}+\frac z2  \right) D_{\nu}(z)=\nu D_{\nu-1}(z),     \eqno(13)
 $$
 (see eqs. (8.2.15-16) in [7]),
 we find
 $$
 {}^+\psi_{1p_v}(x|A_3=-Ex_v)\propto\begin{bmatrix}
(p_1-ip_2)D_{-i\varkappa-1}(\chi_v)\\
e^{-\frac{i\pi}4}\sqrt{\frac{2eE(1-v)}{1+v}}D_{-i\varkappa}(\chi_v)\\
0\\
mD_{-i\varkappa-1}(\chi_v)
    \end{bmatrix}e^{i\vartheta_v},           
                                                           \eqno(14)
$$
  $$
 {}^+\psi_{2p_v}(x|A_3=-Ex_v)\propto
 \begin{bmatrix}
mD_{-i\varkappa-1}(\chi_v)\\
0\\
e^{-\frac{i\pi}4}\sqrt{\frac{2eE(1-v)}{1+v}}D_{-i\varkappa}(\chi_v)\\
-(p_1+p_2)D_{-i\varkappa-1}(\chi_v)
    \end{bmatrix}e^{i\vartheta_v},                    \eqno(15)
$$

From (14-15) for $v=0$ we obtain the usual nonstationary solutions 
$\psi_{p_3}$ with
quantum number $p_3$. On the other hand the solutions $\psi_{p_v}$ can be
 obtained from $\psi_{p_3}$ by a boost along $x_3$  with velocity $-v$.
 Then $\psi_{p_3}$ is multiplied by the matrix [8]
 $$
 \cosh\frac{\varphi}2-\alpha_3\sinh\frac{\varphi}2,\quad \varphi=
 \frac12\ln\frac{1+v}{1-v}.                               \eqno(16)
$$
  In spinor representation this matrix has
 the form (see eq. (4) for $\alpha_3$ in this representation)
 $$
 {\rm diag}(e^{-\frac{\varphi}2}, e^{\frac{\varphi}2},
  e^{\frac{\varphi}2}, e^{-\frac{\varphi}2})=
  \left(\frac{1-v}{1+v}\right)^{-1/4}
{\rm diag}\left(1, \left(\frac{1-v}{1+v}\right)^{1/2},
 \left(\frac{1-v}{1+v}\right)^{1/2}, 1\right).            \eqno(17)
  $$                                                          

  The transition between the standard and spinor representations is made
  with the help of matrix [8] 
  $$
  U=\frac1{\sqrt2}\left(\begin{array}{cc}
1&1\\
1&-1
\end{array}\right). \quad I=\left(\begin{array}{cc}
1&0\\
0&1
\end{array}\right), \quad U^2=1.  \eqno(18)             
  $$
  For $v=1$ we get the solutions with quantum number $p^-=p^0-p_3$. In the
  standard representation we find 
 $$
 {}^+\psi_{rp^-}(x|A_3=-eEx^-)=\frac12[u'_r+\frac{u_r}{\pi^-}]
 \exp\{i[-\frac{p^-x^+}{2}-\frac14eE(x^-)^2-\varkappa\ln z]\},
 $$
  $$
 {}_-\psi_{rp^-}(x|A_3=-eEx^-)=-\frac12[u'_r+\frac{u_r}{\pi^-}]
 \exp\{i[-\frac{p^-x^+}{2}-\frac14eE(x^-)^2-\varkappa\ln(-z)]\}. \eqno(19)
 $$                                                     
 Here
 $$
 z=\frac{\pi^-}{\sqrt{eE}},\quad \pi^-=p^-+eEx^-,\quad x^{\pm}=t\pm x_3,
 \quad r=1,2,
 $$
  $$
  u'_1= \begin{bmatrix}
0\\
1\\
0\\
1
    \end{bmatrix},\quad u'_2=\begin{bmatrix}
1\\
0\\
-1\\
0
    \end{bmatrix},\quad  u_1= \begin{bmatrix}
p_1-ip_2\\
m\\
p_1-ip_2\\
-m
    \end{bmatrix},\quad u_2=\begin{bmatrix}  
m\\
-p_1-ip_2\\
m\\
p_1+ip_2
    \end{bmatrix}.                   \eqno(20)
  $$
  As in the scalar case [6], we have
  $$
  \exp[-i\varkappa\ln z]=e^{-\pi\varkappa}\exp[-i\varkappa\ln(-z)],\quad
  z<0;
  $$
  $$
    \exp[-i\varkappa\ln(-z)]=e^{-\pi\varkappa}\exp[-i\varkappa\ln z
  ],\:
  z>0.     \eqno(21)
  $$
Two other solutions are
$$
{}_+\psi_{rp^-}=\theta(\pi^-)c_{1n}{}^+\psi_{rp^-},\:{}^-\psi_{rp^-}=
\theta(-\pi^-)c_{1n}\:{}_-\psi_{rp^-}, \quad n=(r, p_1, p_2, p^-),
$$
  $$
 \theta(x)=\left\{\begin{array}{cc}
1,\quad x>0\\
0,\quad x<0,
\end{array}\right.
c_{1n}=i\sqrt{\frac{2\pi}{\varkappa}}\frac{\exp[-\frac{\pi\varkappa}2]
}{\Gamma(-i\varkappa)}.                  \eqno(22)
$$

For any $v$ the in- and out-solutions are related by Bogoliubov
transformations
$$
{}_+\psi_{rp_v}=c_{1n}\,{}^+\psi_{rp_v}+c_{2n}\,{}^+\psi_{rp_v},
$$
  $$
{}_-\psi_{rp_v}=-c_{2n}^*\,{}^+\psi_{rp_v}+c_{1n}^*\,{}^-\psi_{rp_v},\eqno(23)
  $$
   were
$$
|c_{1n}|^2+|c_{2n}|^2=1\quad c_{2n}=e^{-\pi\varkappa},
\quad n=(r. p_1, p_2, p_v). \eqno(23a)
$$
 The Bogoliubov coefficients $c_{1n}$, $ c_{2n}$ depend only on $\varkappa$,
 defined in (8).

 As in the scalar case [6], the solutions with quantum number $p_v$ can
  be obtained from the solutions with quantum number $p^-$ by the 
  integral transformation 
$$
\psi_{rp_v}=\int_{-\infty}^{\infty}dp^-N(p_v,p^-)\psi_{rp^-},   \eqno(24)
$$
  $$
N(p_v,p^-)=(2\pi eE(1-v))^{-\frac12}\exp\{i\frac{[p^-(1+v)]^2-4p^-p_v
(1+v)+2p_v^2}{4eE(1-v^2)}\}.                                     \eqno(25)
$$
Thus, using in the right-hand side of (24) the function
 ${}^+\psi_{rp^-}(x|A_3=-Ex^-)$ from (19), we get
  ${}^+\psi_{p_v}(x|A_3=-Ex^-)$ in the left-hand side.
   Changing the potential to $A_3=-Et$, we
  find in the standard representation
 $$
  {}^+\psi_{rp_v}(x|A_3=-Et)=
  \int_{-\infty}^{\infty}dp^-N(p_v,p^-){}^+\psi_{rp^-}(x|A_3=-Et)=
  $$
  $$
 \frac{\exp[-\frac{\pi\varkappa}{4}]}{2\sqrt{eE(1-v)}}
 [u_rD_{-i\varkappa-1}(\chi_v)+
e^{-\frac{i\pi}4}\sqrt{\frac{2eE(1-v)}{1+v}}u'_rD_{-i\varkappa}(\chi_v)]\times
$$
$$
\exp\{i[\frac{\pi}2+\frac{\varkappa}2\ln\frac{1+v}{2(1-v)}+\tilde\vartheta_v]\},
\quad \tilde\vartheta_v=\vartheta_v+\frac{eEvx_3^2}2,\quad
 \chi_v=e^{\frac{i\pi}{4}}T_v.      \eqno(26)
  $$
Similarly, we find
$$
  {}_+\psi_{rp_v}(x|A_3=-Et)=
  \int_{-\infty}^{\infty}dp^-N(p_v,p^-){}_+\psi_{rp^-}(x|A_3=-Et)=
  $$
  $$
 \frac{\exp[-\frac{\pi\varkappa}{4}]}{2\sqrt{eE\varkappa(1-v)}}
 [u_rD_{i\varkappa}(-\chi^*_v)+
e^{-\frac{i\pi}4}\sqrt{\frac{2eE(1-v)}{1+v}}
\varkappa u'_rD_{i\varkappa-1}(-\chi^*_v)]
\times
$$
$$
\exp\{i[\frac{\pi}2+\frac{\varkappa}2\ln\frac{1+v}{2(1-v)}+\tilde\vartheta_v]\},
\quad  \chi^*_v=e^{-i\frac{\pi}4}T_v.      \eqno(27)
  $$
 We note that ${}^-\psi_{p_v}$  $({}_-\psi_{p_v})$  can be obtained from
${}_+\psi_{p_v}$  $({}^+\psi_{p_v})$  by substitutions $T_v\to-T_v$,
 $u'_r\to-u'_r$.

 To compare (26) with (14-15), we go in (26) to the spinor  representation.
 From (18) and (20) we have
 $$
 Uu'_1=\sqrt2\begin{bmatrix}
0\\
1\\
0\\
0
    \end{bmatrix},\: Uu'_2=\sqrt2\begin{bmatrix}
0\\
0\\
1\\
0
    \end{bmatrix},\: Uu_1=\sqrt2\begin{bmatrix}
p_1-ip_2\\
0\\
0\\
m
\end{bmatrix},\: Uu_2=\sqrt2\begin{bmatrix}
m\\
0\\
0\\
-p_1-ip_2
    \end{bmatrix}.         \eqno(28)  
 $$
Now we see that (26) indeed agrees with (14).

Thus, the transformation (24) relates the solutions $\psi_{rp^-}$
to $\psi_{p_v}$. For $v=0$  the latter solutions become solutions
 $\psi_{rp_3}$ and they coincide up to an unimportant phase factor
 $\exp[i\frac{\pi}2-\frac{i\varkappa}2\ln2]$ with the normalized solutions
 obtained in [9], see also [4].  This justifies the normalization factors 
 and the classification of $\psi_{rp^-}$  solutions in (19).

 We see that the classification in  $\psi_{rp_3}$  (which is beyond doubt)
 determines the classification in $\psi_{rp_v}$ and $\psi_{rp^-}$.
  In the next Section we shall see that the classification in $\psi_{rp^-}$
in its turn determines the classification in $\psi_{rp_s}$ and $\psi_{rp^0}$.

\section{ Stationary solutions}

These solutions appear naturally when we use instead of the vector-potential (2)
the potential
 $$ 
 A^0=-Ex_s,\quad x_s=x_3-st,\quad 0\le s<1.   \eqno(29)
 $$
 The solutions of the squared Dirac equation (3) can be obtained from the
 corresponding solutions of the Klein-Gordon equation (see eq. (59) in 
 [6]) by the same substitutions (7). For positive-frequency in-solution
 we find
 $$
{}_+F=\exp[i\vartheta_s]{\rm diag}(\varphi_1, \varphi_2, \varphi_2,
 \varphi_1), \: \varphi_1=D_{i\varkappa}(\chi_s),\:
\varphi_2=D_{i\varkappa-1}(\chi_s),\: \chi_s=e^{\frac{i\pi}4}Z_s,  
     \eqno(30)
$$                                  
  $$
  Z_s=\sqrt{\frac{2}{eE(1-s^2)}}(eEx_s-p_s),\:p_s=p^0-sp_3,\:
\vartheta_s=-\frac{p_s(t-sx_3)}{1-s^2}-\frac{eEsx_s^2}{2(1-s^2)}. \eqno(31)
  $$
 The solutions of the Dirac equation are given by the columns of the 
 matrix in eq. (10), where now
 $$
\Pi^0=i\frac{\partial}{\partial t}-eEx_s,
\quad \Pi_3=-i\frac{\partial}{\partial x_3}.    
$$
From here and (30) we get
$$
(\Pi^0-\Pi_3)e^{i\vartheta_s}\varphi(\chi_s)=-e^{i\vartheta_s-\frac{i\pi}4}
\sqrt{\frac{2eE(1-s)}{1+s}}\left(\frac{\partial}{\partial\chi_s}
+\frac{\chi_s}{2} \right)\varphi(\chi_s).           \eqno(32)
$$
Using again the fourth and the thirst columns in (9),  we obtain
in the spinor representation
$$
 {}_+\psi_{1p_s}(x|A^0=-Ex_s)\propto\begin{bmatrix}
(p_1-ip_2)D_{i\varkappa}(\chi_s)\\
-e^{\frac{i\pi}4}\sqrt{\frac{2eE(1-s)}{1+s}}\varkappa D_{i\varkappa-1}(\chi_s)\\
0\\
mD_{i\varkappa}(\chi_s)
    \end{bmatrix}e^{i\vartheta_s},           
                                                           \eqno(33)
$$
  $$
 {}_+\psi_{2p_s}(x|A^0=-Ex_s)\propto
 \begin{bmatrix}
mD_{i\varkappa}(\chi_s)\\
0\\
-e^{\frac{i\pi}4}\sqrt{\frac{2eE(1-s)}{1+s}}\varkappa D_{i\varkappa-1}(\chi_s)\\
-(p_1+p_2)D_{i\varkappa}(\chi_s)
    \end{bmatrix}e^{i\vartheta_s}.                    \eqno(34)
$$

The normalized solutions $\psi_{rp_s}$ can be obtained from $\psi_{rp^-}$
in (19) by the same integral transformation as in the scalar case [6]
$$
\psi_{rp_s}=\int_{-\infty}^{\infty}dp^-S(p_s,p^-)\psi_{rp^-}, \eqno(35)
$$
  $$
S(p_s,p^-)=(2\pi eE(1-s))^{-\frac12}\exp\{-i\frac{[p^-(1+s)]^2-4p^-p_s
(1+s)+2p_s^2}{4eE(1-s^2)}\},                                     \eqno(36)
$$
For the potential
$$
{\cal A}_{\mu}=\delta_{\mu0}Ex_3, \quad {\cal A}^0=-{\cal A}_0. \eqno(37)
$$
we find in standard representation
$$
 {}^+\psi_{rp_s}(x|{\cal A})=
 \frac{\exp[-\frac{3\pi\varkappa}{4}]}{2\sqrt{eE(1-s)}}
 [u_rD_{-i\varkappa-1}(\chi^*_s)-
e^{\frac{i\pi}4}\sqrt{\frac{2eE(1-s)}{1+s}}u'_rD_{-i\varkappa}(\chi^*_s)]\times
$$
$$
\exp\{i[\frac{\pi}2+\frac{\varkappa}2\ln\frac{1+s}{2(1-s)}+\tilde\vartheta_s]\},
\quad \tilde\vartheta_s=\vartheta_s-\frac{eEst^2}2,\:\chi^*_s=
e^{-\frac{i\pi}4}Z_s,      \eqno(38)
  $$
$$
 {}_+\psi_{rp_s}(x|{\cal A})=
 \frac{\exp[-\frac{3\pi\varkappa}{4}]}{2\sqrt{eE\varkappa(1-s)}}
 [u_rD_{i\varkappa}(\chi_s)-
e^{\frac{i\pi}4}\sqrt{\frac{2eE(1-s)}{1+s}}\varkappa u'_rD_{i\varkappa-1}
(\chi_s)]\times
$$
$$
\exp\{i[\frac{\pi}2+\frac{\varkappa}2\ln\frac{1+s}{2(1-s)}+\tilde\vartheta_s]\}.
     \eqno(39)
  $$
  The solutions ${}^-\psi_{rp_s}$  (${}_-\psi_{rp_s}$) can be obtained 
  from ${}_+\psi_{rp_s}$  (${}^+\psi_{rp_s}$) by substitutions
  $Z_s\to-Z_s$,  $u'_r\to-u'_r$. The solutions $\psi_{rp_s}$ satisfy 
  the same Bogoliubov transformations (23) as $\psi_{rp_v}$.

For $s=0$ we get the usual stationary solutions with the quantum number
$p^0$. Their normalization factor can be checked independently
 (up to an inessential
 phase factor)  using the relation between the normalization 
 on unit current and the usual normalization.
  This relation for the 
 scalar particle was established in [6]. The result remains valid
 (up to a sign for negative-frequency states) for the
 spinor particle as well. To apply this result, we first relate our
 classification to the classification according to the sign of 
 momentum in asymptotic regions  $x_3\to\pm\infty$. (In this classification
  the sign  of particle
  momentum along $x_3$ plays the role of sign of energy [4].) To distinguish 
  this classification from the previous one, we denote the solutions
  in this classification as
  $f_n$,( $n$ is the set of quantum numbers) and assume that they are 
  normalized on a
  unit current (up to a sign) along $x_3$. The correspondence 
   between functions in this
  two classifications is
  given by
  $$
  {}^-\psi_n\propto{}_+f_n,\:  {}_-\psi_n\propto{}_-f_n,\:
  {}^+\psi_n\propto{}^+f_n,\:  {}_+\psi_n\propto{}^-f_n.     \eqno(40)
  $$
  Here ${}_+f_n$   (${}_-f_n$) is the solution with only one wave with 
  the positive (negative) momentum along $x_3$  for $x_3\to-\infty$.
  Similarly, for  $x_3\to\infty$  the sign of momentum is indicated 
  in the upper position.

  In terms of $f_n$ eq.(23) can be rewritten as 
   $$
 {}_+f_n=c'_{1n}{}^+f_n+c'_{2n}{}^-f_n,\: 
 {}^+f_n=-c'{}^*_{1n}{}_+f_n-c'_{2n}{}_-f_n;
 $$
   $$
  |c'_{2n}|^2-|c'_{1n}|^2=1,\; c'_{1n}=-\frac{c_{1n}}{c_{2n}},\:
  c'_{2n}=\frac1{c^*_{2n}}                  \eqno(41)
  $$
 (see eqs. (5.17), (5.18) in [4]; $\psi_n$ there are
  $f_n$ here).
 Now the above-mentioned relation between usual normalization 
  and normalization on the unit current can 
 be written in the form
 $$
 \int dx_3J^0({}_+f_{rp'{}^0},{}_+f_{rp^0})
 =|c'_{2p}|^22\pi\delta(p'{}^0-p^0).\eqno(42)
$$
 So, to obtain the normalization on $2\pi\delta(p'{}^0-p^0)$, we have to divide 
 the solution $f_n$ by $|c'_{2n}|$. To normalize $f_n$ on unit current, we use
 as $f_n$ the expression (33) or (34) in
 $$
 J_3=f^*_n\alpha_3 f_n.  \eqno(43)
 $$
 Then we use the relation
 $$
 D_{-i\varkappa}(y)D_{i\varkappa}(iy)-\varkappa D_{i\varkappa-1}(iy)
 D_{-i\varkappa-1}(y)=e^{-\frac{\pi\varkappa}2},    \eqno(44)
 $$                                                              
 which follows from (see eq. (8.2.11) in [7])
 $$
D_{\nu}(y)\frac d{dy}D_{-\nu-1}(iy)-D_{-\nu-1}(iy)\frac d{dy}D_{\nu}(y)=
-ie^{-i\frac{\pi\nu}2}                                        \eqno(45)
$$
and relations (13). In this way we find in the spinor representation, 
where $\alpha_3$ is given in (4), the normalization factor
$$
|N_{rp^0}|=\frac{e^{\frac{\pi\varkappa}4}}{\sqrt{m^2+p_1^2+p_2^2}}
\frac1{|c'_{2n}|}=\frac{e^{-\frac{3\pi\varkappa}4}}{\sqrt{2eE\varkappa}}.
                                                                \eqno(46)
$$                                                                       
This is in agreement with (39) for $s=0$ (and with the expression for
${}^-\psi_{p^0}$ obtainable from ${}_+\psi_{p^0}$ as indicated below eq.(39)),
 if we go there from standard 
to spinor 
representation with the help of (28).
 \section{Concluding remarks}

In this paper we have completed the study of equivalent sets of solutions
 of the Klein-Gordon and Dirac equations with constant electric field
 initiated in [10, 11].
 We find that in the spinor case everything is just as  expected 
 from the scalar case, considered in [6].

 \section{Acknowledgments}
This work was supported in part by the Russian Foundation for Basic Research
(projects no 00-15-96566 and 02-02-16944).
                             
     \section*{References}

\begin{enumerate}
\item
 A.Hansen, F.Ravndal, Physica Scripta, {\bf 23}, 1036 (1981).      \\
\bibitem{2}
 W.Greiner, B.M\"uller, J.Rafelski, {\sl Quantum Electrodynamics of Strong
  Field}, Springer-Verlag (1985). \\
\bibitem{3}
 A.Calogeracos, N.Dombey, Contemp. Phys. {\bf 40}, 313 (1999).         \\
\bibitem{4}
 A.I.Nikishov, Tr. Fiz. Inst. Akad. Nauk SSSR {\bf 111}, 152 (1979)  ;\\
J. Sov. Laser Res. {\bf 6}, 619 (1985).  \\
\bibitem{5}
A.I.Nikishov, hep-th/0111137; Problems of Atomic Science and Technology,
2001, N6, p. 103.\\  
\bibitem{6}
A.I.Nikishov, hep-th/0202024\\  
\bibitem{7} 
 {\sl Higher Transcendental Functions, Vol. 2 (Bateman Manuscript Project
     )},\\
     Ed. by  A.Erd\'elyi (McGraw-Hill, New York, 1953; Nauka, Moscow,
    1980; Pergamon, Oxford, 1982). \\
\bibitem{8}
     L.H.Ryder, {\sl Quantum Field Theory}
(Cambridge University Press,  1985)\\
\bibitem{9}
A.I.Nikishov, Zh. Eksp. Teor. Fiz. {\bf 57}, 1210 (1969); 
Sov. Phys.-JETP {\bf 30}, 660 (1970).\\
\bibitem{10}
 N.B.Narozhny and A.I.Nikishov, Teor.  Mat. Fiz. {\bf 26}, 16 (1976).  \\
\bibitem{11}
 N.B.Narozhny and A.I.Nikishov, Tr. Fiz. Inst. Akad. Nauk SSSR
 {\bf 168}, 175 (1985); \\
 in {\sl Issues in Intensive-Field Quantum Electrodynamics}, Ed. by
 V.L.Ginzburg (Nova Science, Commack, 1987).   \\
\end{enumerate}
\end{document}